\documentstyle[prd,aps,psfig]{revtex}
\newcommand{\be}{\begin{equation}} 
\newcommand{\ee}{\end{equation}}
\newcommand{\ba}{\begin{eqnarray}} 
\newcommand{\ea}{\end{eqnarray}}
\newcommand{\no}{\nonumber \\} 
\newcommand{\bb}{\bibitem}

\newcommand{\mbh}{\mbox{$M_{BH}$}}

\newcommand{\mi}{\mbox{$M_{i}$}}
\newcommand{\mhi}{\mbox{$M_{Hi}$}}
\newcommand{\mh}{\mbox{$M_{H}$}}

\newcommand{\obh}{\mbox{$\Omega_{BH}$}}

\newcommand{\s}{\mbox{$\rm{sec}$}} 
\newcommand{\g}{\mbox{$\rm{g}$}}
\newcommand{\mev}{\mbox{$\rm{MeV}$}}
\newcommand{\gev}{\mbox{$\rm{GeV}$}}
\newcommand{\mstar}{\mbox{$\rm{M_{\odot}}$}}
\newcommand{\delc}{\mbox{$\delta_{c}$}}
\newcommand{\delhc}{\mbox{$\delta^{c}_{H}$}}

\newcommand{\delh}{\mbox{$\delta_{H}$}}
\newcommand{\deli}{\mbox{$\delta_{i}$}}

\newcommand{\simh}{\mbox{$\sigma_{H}$}}

\newcommand{\mpl}{\mbox{$M_{Pl}$}}
\newcommand{\lsim}{\lesssim}
\newcommand{\gsim}{\gtrsim}

\begin{document} 
\draft 
\twocolumn[\hsize\textwidth\columnwidth\hsize\csname@twocolumnfalse\endcsname 
\title{Primordial black holes under the double inflationary power spectrum} 
\author{Hee Il Kim} 
\address
{Basic Science Institute and Department of Physics, Sogang University,
121-742, Seoul, Korea} 
\maketitle
\begin{abstract} 
Recently, it has been shown that the primordial black holes (PBHs) produced by
near critical collapse in the expanding universe have
a scaling mass relation similar to that of black holes produced in
asymptotically flat spacetime. Distinct from PBHs formed with
mass about the horizon mass (Type I), the PBHs with the scaling 
relation (Type II) can be created with a range of masses at a given formation
time. In general, only the case in which the PBH formation is
concentrated at one epoch has been considered. However, it is
expected that PBH formation is possible over a broad range of
epochs if the density fluctuation
has a rather large amplitude and smooth scale dependence. In this 
paper, we study
the PBH formation for both types assuming the power spectrum of double
inflationary models in which the small scale fluctuations could have
large 
amplitudes independent of the CMBR anisotropy. The mass spectrum of
Type II PBHs is newly constructed without limiting the PBH formation
period. The double inflationary power spectrum is assumed to be of double
simple power-law which are smoothly connected. Under the assumed power
spectrum, the accumulation
of small PBHs formed at later times is important and the mass range is
significantly broadened for both Types. 
The PBH mass spectra are far smoother than the observed MACHO spectrum due to
our assumption of a smooth spectrum. In order to fit the observation, a more
spiky spectrum is required.
\end{abstract}

\narrowtext

]

\newpage 
\section{Introduction} 
The density fluctuations give an interesting way to
form primordial black holes (PBHs) in the early universe. 
If the fluctuation amplitude of an
overdense region is of order unity when the fluctuation enters into
the cosmological horizon, the overdense region can evolve into a black
hole. Early studies found that the PBHs so formed have mass about the
horizon mass at the formation time (Type I) \cite{zel,car75}. The PBHs
can cover the complete mass range of black holes considered nowadays:
from a mass of about the Planck mass $\simeq 2 \times
10^{-5} \g$ to the mass of a black hole in the galactic bulge $\sim
10^{8} M_{\odot}$. The Hawking evaporation \cite{haw} of PBHs with mass 
$\lsim 5 \times 10^{14}\g$ can
affect many early universe phenomena. In particular, PBHs with initial 
mass $\simeq 5 \times 10^{14}\g$ would
presently be at the final stage of their evaporation and could be
observed through their energetic particle emission. 
Although PBHs have not yet been detected, the upper limits on the PBH number 
density have been extensively studied \cite{car76,con}. 
When the fluctuations are normalized to the CMBR anisotropy, PBH formation 
strongly constrains the spectral index of the density fluctuations, $ n\lsim
1.25$ \cite{gre1,mp3}.
PBHs surviving today and the massive relics which may remain at the end of a 
PBH lifetime \cite{mac87} are natural candidates
for dark matter. PBHs with mass $\sim
0.5 \mstar$ are also candidates for the observed massive compact halo objects
(MACHOs) \cite{macho1,macho2}.  It has been proposed that PBHs
with the MACHO mass 
can be produced in certain inflationary models
\cite{jun1,kaw,jun2,juna} or at the vanishing of the sound velocity at the
cosmological quark-hadron phase transition \cite{jed}.
If the MACHO
PBHs form coalescing binaries, they could be detected by gravitational wave interferometers \cite{nak}. 

On the other hand, critical phenomena have been exhibited in the
gravitational collapse which produces a black hole. This was
 originally found by Choptuik in the numerical study
of the gravitational collapse of massless scalar fields \cite{cho}. 
Especially, he
found that for one parameter family of field configurations with
controlling parameter $p$, the black hole mass scales as
$(p-p_c)^{\gamma_s}$ near and above the critical parameter $p_c$ with a
universal exponent $\gamma_s \approx 0.37$. 
It has been shown subsequently that similar mass
relations hold for other situations, such as gravitational waves
 \cite{abr} and radiation fluid \cite{eva}. 
Interestingly, all the studies have found that
$\gamma_s \approx 0.37$. The analytic explanation has been 
given by perturbative analysis \cite{koi}. 

Cosmological application of the critical phenomena was recently
noticed by Niemeyer and Jedamzik in the study of the gravitational
collapse of an overdense region in the expanding universe
\cite{nie1,nie2}. They found
that if the horizon crossing amplitude $\delh$ of the overdense region
is near and above the critical amplitude $\delhc \approx 0.7$, 
then the black hole mass
scales similarly, with $\gamma_s \approx 0.36$ (Type II PBHs)
\cite{nie2}. 
The scaling relation can induce great changes in
the mass spectrum because PBHs can form with a range of masses at a
given time, while Type I PBHs have an initial mass of about the
horizon mass at a given time. The mass spectrum of Type II PBHs has
been studied for the case in which the PBHs form over a very short
period \cite{nie1,gre2,jun3}, such as occurs for a blue-shifted density
perturbation spectrum or the spiky spectrum associated with MACHO PBHs. Based
 on these mass spectra, the upper limits on the PBH number density 
and the spectral index have been revised \cite{jun3,gre3}.

With normalization of the density fluctuation 
spectrum to the COBE CMBR anisotropy measurements,
the PBHs can only form significantly under a simple power-law spectrum
 of the density fluctuations if the density fluctuation is largely 
blue-shifted with a fine-tuned spectral index
\cite{gre1,mp3,car94,mp2}. In this case, the PBH formation is
concentrated at the initial time $t_{i}$ when the fluctuation
develops. 
Similarly, a narrow spiky fluctuation spectrum peaked
at the MACHO scale is required for some MACHO PBH models \cite{jun1,jun2,juna}. 
Hence, it has been usually
assumed that PBHs are formed at only one epoch or with only one
fluctuation scale. For Type
I PBHs this assumption leads to the 
 $\delta$-function type mass spectrum \cite{car76}. 
For Type II PBHs, even if they are formed at one epoch, 
the PBHs can have a range of masses because of the scaling mass relation
 \cite{nie1,gre2,jun3}. 

With rather large fluctuation amplitudes and a smooth power
spectrum, the PBH formation is possible over a range of
epochs (or fluctuation scales) 
and the accumulation of PBHs formed at different times should
be considered. However, even in the models with large amplitude and
smooth scale dependence \cite{kaw},
 the accumulation effect has previously been neglected and the mass
 spectrum obtained by simply taking a short duration of PBH
 formation. With an assumption of large amplitude on small scales, we
 will investigate the accumulation effect in the mass spectrum of PBHs
 obtained based on the Press-Schechter method \cite{pre}. 
For this, we newly construct the mass spectrum which
is more adequate to describe Type II PBHs formed over a range of epochs.
Related work on the mass spectrum was done in Ref. \cite{gre2}
using the excursion set formalism, but that reference investigated 
the validity of the
mass spectrum obtained by limiting the PBH formation epoch.
The large amplitude power spectrum is
natural in double inflationary models. In the double
inflationary models, the small scale fluctuations
 could have large amplitude not constrained by the CMBR
anisotropy, leading to enhanced formation of small PBHs. 
Though there are various double inflationary models, 
we will simply assume a double power-law spectrum.
This kind of power spectrum has been studied in
the double inflationary model of supergravity theory proposed 
to explain the MACHOs as PBHs \cite{kaw}. 
From the mass spectrum of PBHs of both types,
it will be shown that the PBH mass spectrum has difficulty explaining 
the observed MACHO spectrum with the assumed power spectrum of the
density fluctuation.

The rest of the paper is organized as follows. In Sec. II, the PBHs of
both types are reviewed. The mass spectrums are formulated in
Sec. III. PBH formation under the double-inflationary power
spectrum is given in Sec. IV. The paper closes with some concluding
remarks in Sec. V.
\section{PBH formation and its mass}
We only consider the PBH formation in a universe with a
hard equation of state for which the sound velocity $v_s \equiv
\sqrt{\gamma}$  with $0<\gamma \lsim 1$. Although we are interested in 
the radiation-dominated era (when $\gamma=1/3$), we regard $\gamma$ as a
parameter for convenience. For an overdense region in the fluctuated
universe to collapse, the
fluctuation amplitude of the region should be large enough to overcome
the Jeans pressure of the region. Conversely, it should not be so large
that the region is decoupled from the background universe. 
These conditions
can be written for the horizon crossing amplitude $\delh$ as
\be
\delhc \le \delh \le \delta_{H}^{d}~~~.
\ee
In addition, $\delh$ should be larger than the critical value needed
for the trapped surface formation.

For Type I PBHs, it has been found from the study of the evolution of a
spherical overdense region in the expanding universe that the bounds
for the PBH formation, $\delhc$ and $\delta^{d}_{H}$,
 are scale independent and should be of the order
of $\gamma$ \cite{car75}. It has been usually taken that $\delhc=\gamma$ and
$\delta^{d}_{H}=1$. Also, it has been assumed that the trapped
surface forms with these bounds at nearly the horizon
time. Then the PBH mass can be approximated,
 neglecting the slight dependence on the fluctuation amplitude, as
\cite{car75,mp3} 
\ba
\mbh(t)=&&\gamma^{3/2}
M_{H}=\gamma^{3/2}M_{Hi}\biggr(\frac{t}{t_{i}}\biggr)
 \no
        =&&
\gamma^{3\gamma/(1+3\gamma)}M^{(1+\gamma)/(1+3\gamma)}
M^{2\gamma/(1+3\gamma)}_{Hi}
\ea
where $M_{H}$ is the horizon mass, the subscript `$i$' represents
the quantity at
$t_{i}$, the time when the density fluctuation develops, and $M
\propto M^{3/2}_{H}$ is the mass contained in the overdense region
with comoving wavenumber $k$ at $t_{i}$. Note that the mass of PBHs
formed at a given time is fixed as given in Eq. (2). So, all the
quantities, $\mbh, M_{H}, M$, and $k$ have one-to-one correspondences. 
Larger PBHs can form only if the larger fluctuations cross
the horizon at later times.

Distinct from Type I PBHs, Niemeyer and Jedamzik investigated the PBH
formation by numerically simulating the evolution of an overdense region in the
expanding universe \cite{nie1,nie2}. Applying the three different
configurations of the overdense region, they found that the critical phenomena
also exist in
the PBH formation and the resulting PBH mass scales as
\be
\mbh=kM_H (\delh-\delhc)^{\gamma_{s}}
\ee
where $k \sim 3$, $\delhc \approx 0.7$, and 
the critical exponent, $\gamma_s \approx 0.36$, is similar to that
obtained in the asymptotically flat cases \cite{nie2}. 
In this case, the
PBH formation mass is not unique at a given time any more but covers a range of
masses, in principle, from zero to infinity. However, we introduce the
Planck mass $\mpl$ as a minimum PBH mass. Though $\delta^{d}_{H}$ has not yet 
been determined for Type II PBHs, we take $\delta^{d}_{H}=1$ as the
upper limit for PBH formation. Actually, $\delta^{d}_{H}$
is not crucial in this work. Then the possible mass range for a given
time $t$ is from $\mpl$ to $kM_{H}(t) (\delta^{d}_{H}-\delhc)^{\gamma_{s}}
\sim 2M_{H}(t)$. Because the critical amplitude
$\delhc=0.7$ is larger than that taken for Type I PBHs 
$\delhc=1/3$, one may suspect that Type I PBHs always form more abundantly 
than Type II PBHs. However this is not so because the mass shift
to smaller PBHs should be considered for Type II PBHs. The details are given in
Sec. IV.
\section{mass spectrum of PBHs}
The mass spectrum of PBHs can be constructed based on the
Press-Shechter method \cite{pre}. 
To do this we start from the description of an
overdense region at $t_{i}$. We only consider Gaussian 
fluctuations.\footnote{The PBH formation from non-Gaussian fluctuations 
have been studied in \cite{bul,iva,jun2}} For 
an overdense region with mass scale $M$ and size $R$, the smoothed density
field $\delta_{M}$ is defined by
\be 
\delta_{M}({\bf{x}})=\int d^{3}{\bf{y}}
\delta({\bf{x}}+{\bf{y}})W_{M}({\bf{y}})
\ee 
where
$\delta({\bf{x}}) \equiv
(\rho({\bf{x}})-\bar{\rho})/\bar{\rho}$, 
$\bar{\rho}$
is the background energy density of the universe, and $W_{M}({\bf{x}})$
is the smoothing window function of scale $M$. The dispersion
$\sigma_{M}$, 
the standard deviation of the density contrast of the regions with
$M$, is given by 
\be \sigma_{M}^{2}=\frac{1}{V^{2}_{W}}\langle
\delta_{M}^{2}({\bf{x}}) \rangle = \frac{1}{V^{2}_{W}}\int
\frac{d^{3}{\bf{k}}}{(2\pi)^{3}}|\delta_{{\bf{k}}}|^{2}W_{{\bf{k}}}^{2}(M)
\ee 
where $V_{W} \sim R^{3}$ denotes the effective volume filtered by
$W_{M}$, and $\delta_{{\bf{k}}}$ and $W_{{\bf{k}}}$ are the Fourier
transforms of $\delta(\bf{x})$ and $W_{R}(\bf{x})$, respectively.

For Gaussian fluctuations, the probability that the region of size $R$ has density
contrast in the range of $(\deli+d\deli, \deli)$ is 
\be P(M,
\deli)d\delta=\frac{1}{\sqrt{2\pi}\sigma_{M}}\exp
\biggr(-\frac{\deli^{2}} {2\sigma^{2}_{M}}\biggr) d\deli~~~.  
\ee
The fluctuation amplitudes at $t_{i}$ are related to the amplitudes at
the horizon crossing time as follows,
\be
\delta_i =
\biggr(\frac{M}{M_{Hi}}\biggr)^{-\frac{2}{3}}\delh,~~~\sigma_{M}=\biggr(\frac{M}{M_{Hi}}\biggr)^{-\frac{2}{3}}\simh~~~.
\ee

Since the smoothed density field under the window with size $R$
sees only the structures larger than $R$,
\be
F(M,\delc)=\int^{\infty}_{\delc} P(M, \delta_i )d\delta_i
\ee
can be interpreted as the fraction of the overdense regions larger
than $R$ and with $\delta_i > \delc$. And the fraction of the regions with
$\delta_i > \delc $ in the range of $(M, M+dM)$ is given by
\be
f(M,\delc)dM =-\frac{\partial F}{\partial M} dM~~~.
\ee

For Type I PBHs, from Eq. (1) and (7), 
the condition for PBH formation can be written for
$\delta_i$ as \cite{car75}
\be \alpha \equiv \gamma
\biggr(\frac{\mi}{M_{Hi}}\biggr)^{-\frac{2}{3}} \le \delta_{i}
\le \biggr(\frac{\mi}{M_{Hi}}\biggr)^{-\frac{2}{3}}
\equiv \beta~~~.
\ee 
From the PBH formation condition, the fraction of overdense
regions which will evolve to PBHs is
$
f_{BH}dM= f(M,\alpha)dM-f(M,\beta)dM \simeq f(M,\alpha)dM
$. Multiplying by $\bar{\rho_{i}}/M$, this fraction can be converted
to the number density.
After a change of variables, the initial mass spectrum of PBHs in
the range of $(\mbh,~~\mbh+d\mbh)$ is given by \cite{mp2}
\ba
n_{BH}&&(\mbh)d\mbh=-\sqrt{\frac{2}{\pi}}\gamma^{\frac{7}{4}}
\frac{\bar{\rho_{i}}}{M_{Hi}}\biggr(\frac{\mbh}{M_{Hi}}\biggr)^{-\frac{3}{2}}
\times \no
&&\biggr[\frac{\mbh}{\sigma^{2}_{H}}\frac{\partial \simh}{\partial
\mbh}-\frac{1}{\simh}\biggr] \exp
\biggr(-\frac{\gamma^{2}}{2\sigma^{2}_{H}}\biggr)\frac{d\mbh}{\mbh}~~~.  
\ea
Here a factor of 2 is included in accordance with the
Press-Schechter prescription for normalization \cite{pre}.
The number density of all the PBHs at time $t$ is then given by
\be
\beta_{BH}(t)=\biggr(\frac{a(t)}{a(t_{i})}\biggr)^{-3} 
\int^{M_{BH2}}_{M_{BH1}} n_{BH}(\mbh)d\mbh
\ee
where $a(t)$ is the cosmic scale factor, $M_{BH2}=\gamma^{3/2} M_{H}(t)$, and 
$M_{BH1}=\mhi$ if $M_{\ast}(t)\simeq 10^{9}(t/1\s)^{1/3}
\g < \mhi$ where $M_{\ast}(t)$ is
the PBH mass whose lifetime is $t$, else $M_{BH1} =M_{\ast}(t)$. 
If PBHs leave stable massive relics at the end of their evaporation lifetimes,
 then $M_{BH1}=M_{Hi}$. 

If the power spectrum  can be approximately treated as a
$\delta$-function type as in the blue-shift perturbation with the CMBR
normalization or the spiky spectrum in some inflationary models, it
is usual to assume that the PBH formation occurs at only one epoch.
In this case, the PBHs form with only one mass for Type I PBHs, and 
the initial number density is approximately expressed by \cite{car76}
\be
\beta_i(\mbh) = \simh \exp
\biggr(-\frac{\gamma^{2}}{2\sigma^{2}_{H}}\biggr)
\ee
For Type II PBHs, even if the power spectrum resembles a $\delta$-function, 
the PBHs can form with a range of masses from about $\mpl$ to $2M_{H}(t)$. 
Their mass spectrum can be obtained from the probability
distribution of $\delta_i$ (or $\delh$), Eq. (6), as follows,
\ba
n_{BH}&&(\mbh)d\mbh=\frac{1}{\sqrt{2\pi} \simh \gamma_{s}}
\frac{\bar{\rho}_{i}}{\mhi}
\biggr(\frac{\mh}{\mhi}\biggr)^{-3/2} \no
\times&& \biggr(\frac{\mbh}{kM_{H}}\biggr)^{1/\gamma_s}
\exp
\biggr(-\frac{\delh^2}{2\simh^2}\biggr)\frac{d\mbh}{\mbh}
\ea
where, from Eq.(3), 
\be
\delh=\delhc+\biggr(\frac{\mbh}{kM_{H}}\biggr)^{1/\gamma_s}~~~.
\ee
The excursion set formalism of \cite{gre2} validated Eq. (14) 
for the blue-shifted fluctuation spectrum with
small CMBR anisotropy and the spiky spectrum for MACHO PBHs
\cite{gre2}. However, the above expressions Eq. (13) and (14) are 
inappropriate to describe the accumulation of PBHs formed at later 
times. In particular, for Type II PBHs, the accumulation of
newly formed PBHs contributes to all masses below $\sim 2M_{H}(t)$, so
that the shape of the mass spectrum is time dependent
 while that of
Type I PBHs, Eq. (11), does not depend on time. For Type I PBHs, the newly
formed PBHs just add to the total number density at masses higher than the 
older PBHs.

One of the main purpose
of this work is to formulate the mass spectrum for Type II PBHs which
describes the accumulation effect more properly.
For this we start from the interpretation of the probability distribution,
Eq.(6). As mentioned before, the smoothed field can see the structures
larger than the window size. So, Eq.(6) can be interpreted as the
fraction of the overdense regions which are larger than the scale $M$
and have amplitude in the range of $(\delta_i , \delta_i + d\delta_i )
$. The contribution from larger scales can be eliminated by
differentiation. Then the fraction of the regions between $(M, M+dM)$ and
$(\delta_i , \delta_i + d\delta_i )$ is given by
\be
\tilde{f}(M,\delta_i )d\delta_i dM = 
-\frac{\partial P(M,\delta_i )}{\partial M} d\delta_i dM~~~.
\ee
Applying a change of variables $(M\rightarrow \mh,~~\deli
\rightarrow \mbh$) and the Press-Schechter prescription,
 we can find the mass spectrum
for the PBHs. The number density of PBHs in the ranges of
$(\mbh,~~\mbh+d\mbh)$ and $(\mh,~~\mh+d\mh)$ can be derived to be
\ba
\tilde{n}_{BH}&&d\mh
d\mbh=-\sqrt{\frac{2}{\pi}}\frac{1}{\gamma_s}
\frac{\bar{\rho}_{i}}{\mhi}
\biggr(\frac{\mh}{\mhi}\biggr)^{-3/2} \no
&&\times\biggr(\frac{\mbh}{k\mh}\biggr)^{1/\gamma_s} 
\biggr(\frac{\delh^2}{\simh^2}-1\biggr)
\biggr[\frac{\mh}{\sigma^{2}_{H}}\frac{\partial \simh}{\partial
\mh}-\frac{1}{\simh}\biggr] \no
&&\times \exp
\biggr(-\frac{\delh^{2}}{2\sigma^{2}_{H}}\biggr)\frac{d\mh}{\mh}\frac{d\mbh}{\mbh}~~~.  
\ea
The mass spectrum at time $t$ is then given by 
\be
n_{BH}(\mbh,t) d\mbh
=\biggr(\int^{M_{H2}}_{M_{H1}} \tilde{n}_{BH}d\mh
\biggr) d\mbh
\ee
where
$M_{H1}=\mbh/[k(\delta^{d}_{H}-\delta^{c}_{H})^{\gamma_s}]$
and $M_{H2}=\mh(t)$. The number density of PBHs at $t$ is
\be
\beta_{BH}(t)=\biggr(\frac{a(t)}{a(t_{i})}\biggr)^{-3} 
\int^{M_{BH2}}_{M_{BH1}} n_{BH}(\mbh)d\mbh
\ee
where $M_{BH2}=k\mh(t)(\delta^{d}_{H}-\delhc)^{\gamma_s}$ and 
$M_{BH1}=\mpl$ if $M_{\ast}(t) < \mpl$, otherwise $M_{BH1} =M_{\ast}(t)$. If
 PBHs leave stable relics, then we can take $M_{BH1}=M_{relic}$. 
\section{PBHs under the double inflationary power spectrum}
With the COBE normalization on the CMBR anisotropy and a simple power-law
spectrum, 
the PBHs can form significantly
 only if the spectrum is strongly blue-shifted. The PBH formation is then 
usually assumed to occur only at one epoch and the accumulation
 of black holes formed at later times
is neglected. 
However if the power spectrum can have larger amplitude, it is expected that 
the PBH formation is possible over a rather broad
period of times. For
practical purposes, one would require the large amplitude to occur on scales
sufficiently small as to be irrelevant to the CMBR anisotropy while the
fluctuation amplitudes on large scales are constrained by the CMBR
anisotropy $\delta_{H} \sim 10^{-5}$. This kind of power spectrum can
be generated by double inflationary models \cite{kaw,jun2,juna,dou,dou1}. In
particular, the double inflation model with one scalar field has been studied
in Refs. \cite{jun2,juna}.
In the double inflationary models, the first inflation gives the density
fluctuation which is responsible for the CMBR anisotropy on large
scales. The second inflation generates the smaller scale
density fluctuations which are irrelevant to the CMBR anisotropy.
If the small scale fluctuations have a rather large amplitude, then the
PBH formation can be enhanced. Phenomenologically, the large amplitude
on small scales in the double inflation model could resolve the problems of the
standard cold dark matter scenario \cite{dou}. It has been also claimed that
MACHO PBHs could be produced in models with large amplitude at the
MACHO scale $\sim 1\mstar$, such as the double inflation from
supergravity theory \cite{kaw} and chaotic new inflationary potential 
\cite{jun2,juna}. The double inflation models with two scalar fields might suffer 
from the initial condition problem if our universe originated from an 
inflationary domain of Planckian density \cite{hod}. 
However, the study on this problem
is not the subject of this work and we assume that the double inflation is 
realized in a proper way.

We consider only the
fluctuation scales larger than $\mhi$, the horizon scale at the reheating time
of the second inflation $(\equiv t_i )$. 
The PBHs from the fluctuations smaller than $\mhi$, if any, will be
diluted away by the second inflation.
 Also, for simplicity, we assume that the power spectrum larger than
$\mhi$ is a combination of two simple power-law spectra which are
given by
\ba
\simh(\mh)=&&\simh_i \biggr(\frac{\mh}{\mhi}\biggr)^{(1-n)/4}~~~~~{\rm{for}}
~~\mh<M^{c}_{H} \no
=&&\sigma_{H0} \biggr(\frac{\mh}{M_{H0}}\biggr)^{(1-n_{obs})/4}
~{\rm{for}}~~\mh \gg M^{c}_{H}
\ea
where the present values $M_{H0} \approx 10^{56}\g$, 
$\sigma_{H0} \simeq 10^{-4}$ \cite{gre1}
 and $n_{obs}=1.2\pm 0.3$ from the
COBE data \cite{gor}, and the breaking scale $M^{c}_{H}$ is 
well below $10^{15} \mstar$ which is the horizon mass at the matter-radiation
equal time \cite{dou}. The two simple power-law spectra are assumed to be
connected in such a way that $\simh$ decreases monotonically to the second 
spectrum for$\mh > M^{c}_{H}$. Also the change of $\simh$ is assumed to be 
smooth satisfying $(M_{H}/\simh)(d\simh/dM_{H}) \ll 1$. If the spectra were 
connected sharply or discontinuously 
at $M^{c}_{H}$, PBHs or other structures with the mass of about $M^{c}_{H}$ 
might be produced significantly. If $M^{c}_{H} \sim 10^{33} \g$, such PBHs
could be related to the MACHOs. However, with the above assumptions,
 we will not consider such PBH formation in this work. 
Also, by assuming $n_{obs} \simeq 1$, the PBHs from the
fluctuation larger than $M^{c}_{H}$ are neglected.

Then the mass spectrum of Type I PBHs with the assumed 
power spectrum is given by
\ba
n_{BH}(\mbh)d\mbh=&&\frac{n+3}{4}\sqrt{\frac{2}{\pi}}\gamma^{\frac{7}{4}}
\frac{\bar{\rho_{i}}}{M_{Hi}}\biggr(\frac{\mbh}{M_{Hi}}\biggr)^{-\frac{3}{2}}
\no \times&&\frac{1}{\simh}\exp
\biggr(-\frac{\gamma^{2}}{2\sigma^{2}_{H}}\biggr)\frac{d\mbh}{\mbh}~~~.  
\ea
Fig. 1 shows the mass spectrums for various parameters. The PBH mass
 starts from $M_{BHi}=\gamma^{3/2}M_{H}(t_{i})$. 
For an $n=1$ scale invariant spectrum, the mass
spectrum has no cutoffs and scales as $n_{BH} \propto \mbh^{-5/2}$. 
For $n<1$ $(n>1)$, there are exponential cutoffs at small (large)
scales. However, as the fluctuation amplitude grows or as $n$ approaches
1, the cutoffs develop more
slowly and the mass range is significantly broadened.

\begin{figure}
\centerline{\psfig{file=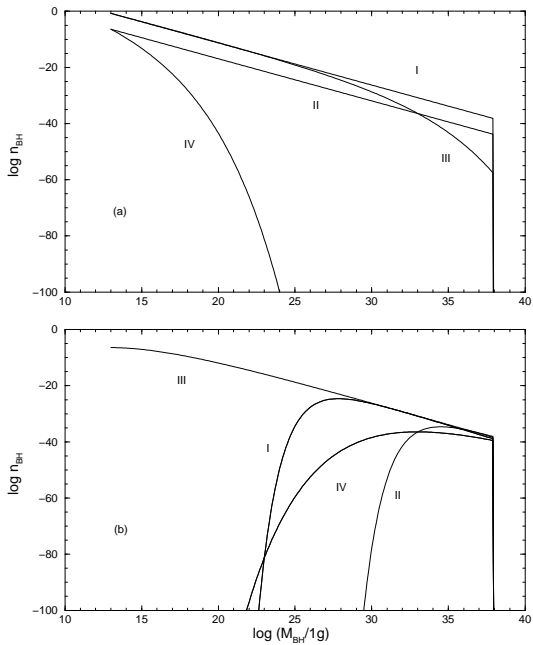,width=3.in}}
\caption{The mass spectrum of Type I PBHs today for
  $T_{RH}=10^{10}\gev$ and $M^{c}_{H}=10^{5}\mstar$; (a) for $n\ge 1$,
  I: $n=1.0$, $\sigma_{Hi}=0.6$, II: $n=1.0$,  
  $\sigma_{Hi}=0.06$, 
III: $n=1.2$, $\sigma_{Hi}=0.6$, and 
IV: $n=1.2$,  $\sigma_{Hi}=0.06$,
(b) for $n < 1$, 
  I: $n=0.4$, $\sigma_{H}(M_{MACHO})=0.6$, II: $n=0.4$,  
  $\sigma_{H}(M_{MACHO})=0.06$, 
III: $n=0.8$,   $\sigma_{H}(M_{MACHO})=0.6$, and  
IV: $n=0.8$,  $\sigma_{H}(M_{MACHO})=0.06$. }
\end{figure}

For Type II PBHs, the mass spectrum is more complicated. 
With the assumed double power spectrum, Eq. (17) becomes
\ba
\tilde{n}_{BH}&&d\mh d\mbh=\frac{n+3}{4}\sqrt{\frac{2}{\pi}}
\frac{1}{\gamma_s}\frac{\bar{\rho}_{i}}{\mhi}
\biggr(\frac{\mh}{\mhi}\biggr)^{-3/2} \no
&&\times \biggr(\frac{\mbh}{k\mh}\biggr)^{1/\gamma_{s}}
\frac{1}{\simh}\biggr(\frac{\delh^2}{\simh^2}-1\biggr) \no
&&\times 
\exp\biggr(-\frac{\delh^{2}}{2\sigma^{2}_{H}}\biggr)\frac{d\mh}{\mh}
\frac{d\mbh}{\mbh}~~~.  
\ea

\begin{figure}
\centerline{\psfig{file=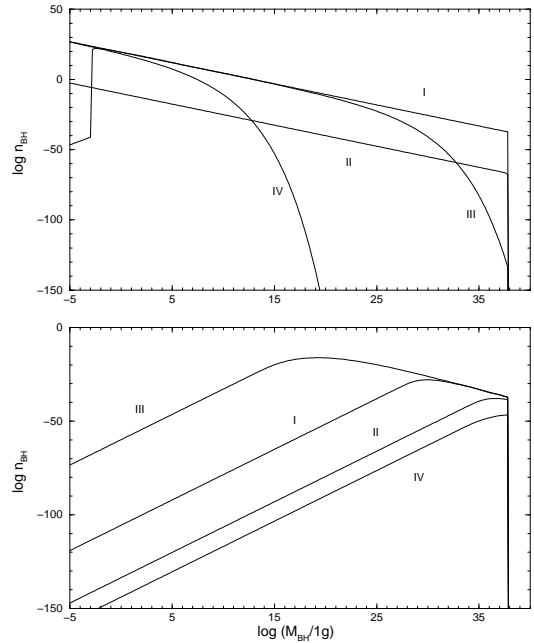,width=3.in}}
\caption{The mass spectrum of Type II PBHs today with the same
  parameter set of Fig. 1.}
\end{figure}

Fig. 2 shows the mass spectrum $n_{BH}(\mbh,t_{0})$. The PBH mass spectrum
extends down to $\mpl$. There is an extra condition for Type II PBHs:
requiring positive number density requires $\delta^{2}_{H} \ge
\sigma^{2}_{H}$. This condition causes a rapid drop in the
mass spectrum at small scales for $n\geq 1$ or 
at large scales for $n<1$. The drop can be seen in III of Fig. 2a. 
The mass range is also
significantly broadened when the fluctuation amplitude is large or $n$
approaches 1.
Since $\delh$ is a function of $\mbh$, the role of the exponential term
$\exp(-\frac{\delta^{2}_{H}}{2\sigma^{2}_{H}})$ is very different to
that of $\exp(-\frac{\gamma^{2}}{2\sigma^{2}_{H}})$ in the Type I mass
spectrum which leads to cutoffs in the mass spectrum. Interestingly, the
mass spectrum for Type II PBHs scales in the same manner as for Type I PBHs,
$n_{BH}\propto M^{-5/2}_{BH}$, in spite of the enhanced formation of small 
PBHs.  This is mainly because 
$\exp(-\frac{\delta^{2}_{H}}{2\sigma^{2}_{H}})$ is a function of
$(\mbh/\mh)$ for $n=1$ and the main contribution to $\int \tilde{n}_{BH} d\mh$
comes from $\mbh \approx \mh$. For $n>1$, the exponential cutoff
develops similarly at large PBH mass and there is an $\mbh^{-5/2}$
decrease at small PBH mass. However for $n<1$ there is no exponential
cutoff at small mass because of the accumulation of small
PBHs formed at later times. Instead, the mass spectrum at small mass
scales as $n_{BH}\propto \mbh^{(1-\gamma_{s})/\gamma_{s}}$.

For both Types, the dependence on the reheating time only
appears in the mass spectrum for $n\geq 1$. The mass
spectrum shifts to small mass as the reheating time decreases. 
Unfortunately, we now have an extra parameter $M^{c}_{H}$. 
The breaking scale $M^{c}_{H}$ must be
determined, in addition to the shape of the power spectrum, to estimate the
number density $\beta_{BH}$ or the density fraction $\obh$ of PBHs. 
This could be done for more concrete models. Although $\beta_{BH}$ and
$\obh$ are not determined here, it is interesting to compare the mass
spectrum of PBHs with that of MACHOs. From the data fit of 8 MACHO
events, the fraction of MACHOs between $M_{MACHO}$ and
$M_{MACHO}+dM_{MACHO}$ is found to be
\be
f_{MACHO}dM_{MACHO} \propto  M^{-3.9}_{MACHO}dM_{MACHO}
\ee
for $M_{MACHO}\ge 0.3\mstar$ \cite{macho2}.
This fraction corresponds to a PBH mass spectrum which is far steeper than 
the spectra for $n<1$ shown
in Fig. 1b and Fig. 2b. In Ref. \cite{kaw}, it was claimed that the
second inflation with an $n<1$ spectrum could produce PBHs peaked at the
MACHO mass. However, they did not consider the accumulation effect. It
seems difficult to generate the current MACHO spectrum with the assumed two 
simple power-law spectra unless the spectra are connected in a proper way near
$M_{MACHO}$. The MACHO spectrum can arise for some $n>1$ spectra,
but the universe would be overclosed by the PBHs.
\section{concluding remarks}
In this paper, we study the formation of PBHs of two different types,
the PBHs with horizon mass (Type I) and the PBHs with scaling mass
relation (Type II), under the double inflationary power spectrum. The
assumed power spectrum is a double simple power-law spectrum such that the
small scale power spectrum gives PBH formation and large scale power
spectrum generates the CMBR anisotropy. With a rather large fluctuation
amplitude, the PBH formation occurs over a range of epochs and it is
important to consider the accumulation of smaller mass PBHs formed at later
times. For both types, the PBH mass ranges are significantly
broadened. As expected, there are cutoffs which develop slowly in the
mass spectrum for Type I PBHs. From the newly constructed mass
spectrum for Type II PBHs, it is found that the mass spectrum scales
in the same manner as Type I PBHs for $n=1$. Due to the enhanced formation of
small PBHs for Type II PBHs, the mass spectrum for $n<1$ does not
develop an exponential cutoff at small mass but scale as
$\mbh^{(1-\gamma_{s})/\gamma_{s}}$. This may give an unique feature of
  PBHs in the history of the universe.

With the assumed power spectrum, the PBH mass spectrum has difficulty
explaining  the current MACHO spectrum. In the present work, however, 
we did not
consider the spiky spectrum \cite{jun1,jun2,juna} which has more rapid scale
dependence. Since the resulting mass spectrum will also have a rapid
scale dependence, one could explain the MACHO spectrum by controlling
the model parameters. However, the Press-Schechter method could not be
 applicable if the $\simh$ grows rapidly as can be seen for a simple power-law
 with $n<-3$. So, our mass spectrum could not be applicable to the spiky 
spectrum. This problematic issue will be studied in detail later.
 
The gamma-rays from evaporating PBHs can generate a diffuse $\gamma$-ray
background (DGB) in the universe \cite{car76,pag,mac91,add}. However, it is
difficult to generate the complete observed DGB in the $\gamma$-ray energy 
range $0.8 \mev \le E \le 100 \gev$ \cite{sre,kap}  with the PBHs from a simple
power-law spectrum \cite{mp3,kri}. 
For $E \gsim 100 \mev$, the PBH $\gamma$-ray spectrum
falls off $E^{-3}$, faster than the observed DGB spectrum, independent of 
the mass spectrum of PBHs \cite{pag,mac91} 
(or $E^{-4}$ if the PBH emission can form a
photosphere around  the PBH \cite{hec}). 
On the other hand, the $\gamma$-ray spectrum
for $E\lsim 100 \mev$ depends on the mass spectrum of PBHs
in the range of $2\times 10^{13}\g \lsim \mbh \lsim 5\times 10^{14}\g$
\cite{mp3}. 
The double inflation model could be tuned to form sufficient PBHs to
explain the observed $E \lsim 100 \mev$ DGB spectrum.

\acknowledgments
The author is grateful to Professor Chul H. Lee and Professor Bum-Hoon
Lee for useful discussion and Dr. Jane H. MacGibbon for reading the 
manuscript. This work was supported in part by the
Korean Ministry of Education (BSRI-98-2414).

\end{document}